%% file: V2MAS_PAPER.tex
\documentclass[conference]{IEEEtran}
\usepackage{amsmath,amssymb,amsfonts}
\usepackage{url}
\usepackage{hyperref}
\usepackage{cite}

\usepackage{listings}
\usepackage{color}

\definecolor{dkgreen}{rgb}{0,0.6,0}
\definecolor{gray}{rgb}{0.5,0.5,0.5}
\definecolor{mauve}{rgb}{0.58,0,0.82}

\usepackage{graphicx}
\usepackage{booktabs}

\usepackage{multirow}

\begin{document}

\title{Testing Self-Organizing Multiagent Systems}  

\author{\IEEEauthorblockN{1\textsuperscript{st} Nathalia Nascimento}
	\IEEEauthorblockA{\textit{Laboratory of Software Engineering (LES)} \\
		\textit{Pontifical Catholic University of Rio de Janeiro (PUC-Rio)}\\
		Rio de Janeiro, Brazil \\
		nnascimento@inf.puc-rio.br}
	\and
		\IEEEauthorblockN{2\textsuperscript{rd} Carlos Lucena}
	\IEEEauthorblockA{\textit{Laboratory of Software Engineering (LES)} \\
		\textit{Pontifical Catholic University of Rio de Janeiro (PUC-Rio)}\\
		Rio de Janeiro, Brazil \\
		lucena@inf.puc-rio.br}
		\and
		\IEEEauthorblockN{3\textsuperscript{nd} Paulo Alencar}
		\IEEEauthorblockA{\textit{David R. Cheriton School of Computer Science} \\
			\textit{University of Waterloo (UW)}\\
			Waterloo, Canada \\
			palencar@cs.uwaterloo.ca}
	\and
	\IEEEauthorblockN{4\textsuperscript{th} Carlos Juliano Viana}
	\IEEEauthorblockA{\textit{Tecgraf Institute} \\
		\textit{Pontifical Catholic University of Rio de Janeiro (PUC-Rio)}\\
		Rio de Janeiro, Brazil \\
		cviana@tecgraf.puc-rio.br}
}
	
\maketitle

\begin{abstract}  
Multiagent Systems (MASs) involve different characteristics, such as autonomy, asynchronous and social features, which make these systems more difficult to understand. Thus, there is a lack of procedures guaranteeing that multiagent systems would behave as desired. Further complicating the situation is the fact that current agent-based approaches may also involve non-deterministic characteristics, such as \emph{learning}, \emph{self-adaptation} and \emph{self-organization} (SASO). Nonetheless, there is a gap in the literature regarding the testing of systems with these features. 
This paper presents a publish-subscribe-based approach to develop test applications that facilitate the process of failure diagnosis in a self-organizing MAS. These tests are able to detect failures at the global behavior of the system or at the local properties of its parts. To illustrate the use of this approach, we developed a self-organizing MAS system based on the context of the Internet of Things (IoT), which simulates a set of smart street lights, and we performed functional ad-hoc tests. The street lights need to interact with each other in order to achieve the global goals of reducing the energy consumption and maintaining the maximum visual comfort in illuminated areas. To achieve these global behaviors, the street lights develop local behaviors automatically through a self-organizing process based on machine learning algorithms.
\end{abstract}

\IEEEpeerreviewmaketitle

%


\textit{keywords: testing; multiagent systems; self-organizing systems; self-organization; internet of things; emergent behavior; machine learning}


\input{samplebody-conf}

\bibliographystyle{IEEEtran}
\bibliography{sigproc}  

\end{document}

%% file: samplebody-conf.tex
  
\section{Introduction} \label{section:introduction}

Multiagent Systems (MASs) involve different characteristics, such as autonomy, asynchronous and social features, which makes these systems more difficult to understand. Thus, there is a lack of procedures guaranteeing that multiagent systems would behave as desired \cite{pvechouvcek2008industrial}. Further complicating the situation is the fact that current agent-based approaches may also involve non-deterministic characteristics, such as \emph{learning} \cite{do2017fiot}, \emph{self-adaptation} and \emph{self-organization} (SASO) \cite{di2008generic}\cite{do2017fiot}. Nonetheless, there is a gap in the literature regarding the inspection of systems with these features. For example, there are very few approaches to evaluate the local interactions between agents in a self-organizing MAS system and the global behavior that emerges
 from these interactions \cite{gardelli2005role} \cite{bernon2006enhancing}. 
One reason is the difficulty of specifying expected results for non-deterministic applications, especially in actual environments.

We consider here the definition of self-organizing systems that has been used by the editors of the IEEE International Conference on Self-Adaptive and Self-Organizing Systems, as follows \cite{SASO2}:
\begin{quote}
	Self-organizing systems work bottom-up. They are composed of a large number of components that interact according to simple and local rules. The global behavior of the system emerges from these local interactions, and it is difficult to deduce properties of the global system by studying only the local properties of its parts. 
\end{quote}

As a self-organizing MAS system enables the emergence of social features based on the behavior of individual agents, to evaluate this kind of system it is necessary to analyze the activities performed by single agents, the interaction among the agents and the behavior that is exhibited by the whole system. 
In \cite{nascimento2017publish}, we presented a preliminary version of a publish-subscribe-based architecture that was implemented\footnote{The source of the test system is available at \\ \href{http://www.inf.puc-rio.br/~nnascimento/MAS-tests.html}{http://www.inf.puc-rio.br/~nnascimento/MAS-tests.html}}  to make feasible the development of multi-level tests based on logging for multiagent systems. By using this platform, it is possible to test the behavior of individual agents and the behavior of group of agents. However, we only showed the usability of our platform by testing a very simple MAS application - a marketplace to buy and sell books on-line. Therefore,  the goal of this paper is to improve this architecture and present an approach that 
makes it possible to diagnose failures in a more complex MAS application, a self-organizing one. 

To test self-organizing applications, our new approach promotes the development of tests separated into two categories: global and local levels (which will be described in Section \ref{sec:testapproach}).

To illustrate and evaluate the use of the proposed approach, we developed a self-organizing MAS application by using the ``Framework for the Internet of Things" (FIoT) \cite{do2017fiot}, which is an agent-based framework for the development of self-adaptive and self-organizing applications based on the Internet of Things (IoT) \cite{iotVision}. 


This experiment is presented in Section \ref{section:app}. The remainder of this paper is organized as follows. Section \ref{sec:relatedwork} presents the related work. Section \ref{sec:background} presents the background, briefly describing the publish-subscribe based architecture to generate tests and the Framework for the Internet of Things (FIoT). Section \ref{sec:testapproach} describes the approach to test self-organizing systems. Section 6 evaluates the test approach, presenting the experimental results and evaluation. The paper ends with some concluding remarks and a discussion about potential future work in Section 7.

	\section{Related Work} \label{sec:relatedwork}

	According to Nguyen et al. (2009) \cite{nguyen2009testing},
	a full testing process of a multiagent system consists of five levels: unit, agent, integration (or group), system (or society) and acceptance. 
 
	Agent test tests the capability of a specific agent to fulfill its goal and to sense and affect the environment. Integration test tests the interaction of agents and the interaction of agents with the environment, ensuring that a group of agents and environmental resources work correctly together \cite{nguyen2009testing}. System test tests the quality properties that the intended system must reach, such as performance \cite{nguyen2009testing}.


	Few approaches for testing the interactions among a group of agents were proposed. In addition, most of them are already only based on the concept of communication sniffer, that is an agent that can intercept messages. For example, Serrano et al. (2012) \cite{serrano2012approach}, which is one of the most recent papers published about testing MASs at the group level, uses ACLAnalyser \cite{botaa2004aclanalyser}, a tool for debugging MAS through the analysis of ACL \cite{fipa} messages. Thus, by using these current test approaches, if an agent exhibits unexpected behavior (failure), a developer has to inspect this failed agent or messages exchanged between agents to find the fault that caused that failure. However, if an agent fails, its failure may be related to a previous and an unexpected behavior of another agent in the environment. This case would be a real problem to some MAS-based approaches, such as that one proposed by Malkomes et al. (2017) \cite{malkomes2017cooperative}, which promotes the development of cooperative agents without using message communication.

	In particular, there is a lack of approaches to assess the emergence process in a self-organizing MAS system \cite{gardelli2005role} \cite{bernon2006enhancing}. Gardelli et al. \cite{gardelli2005role} provides a theoretical system-oriented approach that aims at anticipating design
	decisions at the early MAS design stages.  Bernon et al. \cite{bernon2006enhancing}  provides a simulation-driven approach, which allows the developer to simulate different versions of the application while designing the agents.
	Kaddoum et al. \cite{kaddoum2009characterizing,kaddoum2010criteria} describes some  evaluation criteria that are required to analyze self-* systems. Accordingly, designers should consider some questions to validate the well-functioning of the system and of the self-*mechanism, such as ``is the system able to solve the problem for
	which it is conceived?" and ``is the system able to self-adapt in an efficient way?". In order to investigate these questions, the authors introduce some performance and robustness metrics, such as time (e.g. the number of steps needed by
	agents to reach the solution), the quality of solution (i.e. functional adequacy of the designed system) and time for adaptation.  
	
	
	\section{Background} \label{sec:background}
	
	\subsection{FIoT: A Framework for Internet of Things} \label{sub:FIoT}
	
	
	The Framework for the Internet of Things (FIoT) \cite{do2017fiot} is an agent-based software framework \cite{do2017fiot} to generate different kinds of applications for IoT. It is based on MAS and artificial intelligence paradigms such as neural networks and evolutionary algorithms. 
	
	The main role of FIoT is to produce MAS-based applications with decentralized, autonomous, self-organizing features. Basically, it supports the development of three types of agents: (i) Manager Agents; (ii) Adaptive Agents; and (iii) Observer Agents. The primary role of the Manager Agent is to detect new things that are trying to connect to the system and make that connection. Adaptive Agents control things at the scenario and must execute three key activities in sequence namely: (i) collect data from the thing; (ii) make decisions; and (iii) take actions. The Observer Agent examines the environment to determine if the system is meeting its global goals. See more details about these agents in \cite{do2017fiot} and \cite{nathalia:mestrado:15}.
	
\subsection{Designing Self-Organizing MAS through Neuroevolution}
\label{sub:evolution}
	
	Evolutionary algorithms, such as genetic algorithm, is a well known approach to develop self-organizing multiagent systems \cite{trianni2011engineering}. It allows the emergence of features that were not defined at design-time, such as a communication system \cite{floreano2007evolutionary}. In short, the genetic algorithm is a population-based search algorithm, in which each individual is a solution in a problem space \cite{tanese1989distributed}. The individuals are evaluated by using a fitness function, and the  fittest individuals are selected to produce offspring of the next generation. 
	
	Nolfi et al. \cite{Nolfi2016} describe some experiments where the behavior of agents is  autonomously configured through a neuro-evolutionary algorithm. Each agent uses an artificial neural network to sense the environment and behave accordingly. To optimize their neural networks, finding the fittest configuration (e.g synaptic weights and neural architecture), Nolfi et al. \cite{Nolfi2016} propose a genetic algorithm. Therefore, each individual of the genetic algorithm population represents a configuration of the agent's neural network. In such case, each gene of an individual may represent the strength of a connection between two neurons.

	The interested reader may consult more extensive papers \cite{trianni2011engineering} and \cite{do2017fiot}.

	\subsection{Failure Diagnosis with Logs Containing  Meta-Information Annotations} \label{subsection:logs}
		Ara\'ujo and Staa \cite{thiagopuc} investigated common approaches for testing distributed systems. According to these authors, there are several approaches that perform diagnosis based on log collection. Nonetheless, they have some limitations, such as the need of (i) organizing logs in a centralized architecture and in an adequate time order; (ii) providing visualization tools to assist manual inspection; and (iii) increasing the log details in order to enable the tool to also diagnose the application's logic. Therefore, they presented a diagnosing mechanism based on logs of events annotated with contextual information, allowing a specialized visualization tool to filter them according to the maintainer's needs. 
		
		In their approach, each logged event records a set of properties, represented as tags. A tag is a key-value pair where the value is optional. Every event must contain a basic set of tags which are: 1) \emph{timestamp}: used to sort all events into a single timeline;
		2)
		\emph{message}: a description of the event;
		3)\emph{request id}: used to identify the type of event;
		4) \emph{device}: used to identify the device that originated the event;
		5) \emph{module}: the module that triggered the notification; and
		6) \emph{line}: the line of code where the notification command was inserted.
	
		\subsection{RabbitMQ: Publish-Subscribe Platform}\label{subsection:rabbit}
		RabbitMQ \cite{rabbitsite} is a message-oriented middleware, which generates asynchronous, decoupling applications by separating sending and receiving data through a client and scalable server architecture. It can be easily integrated into an application to operate as a common platform to send and receive messages, maintaining messages in a safe place to live until received. RabbitMQ is a multi-platform that may be deployed in Java, C, Python, and many other programming languages. It can also be deployed in a cloud infrastructure.

		By using RabbitMQ, it is possible to build a logging system based on the publish-subscribe architecture. The publisher is able to distribute log messages to many receivers, while the consumers have the possibility of selectively receiving the logs. Publisher and consumers communicate through queues. Each queue has a particular routing key that is a list of words, delimited by dots. There can be as many words in the routing key as you like, up to the limit of 255 bytes. These words can be anything, but usually they specify some features connected to the message. For example, if a developer specifies that a log message must meet the pattern ``(month).(day).(deviceId).(typeLog)", the valid routing keys would be ``november.11.device01.error" and ``november.
		15.device01.info"  \cite{rabbitsite}.
		
		Therefore, a message sent with a particular routing key will be delivered to all the queues that are bound with a matching binding key. However there are two important special cases for binding keys \cite{rabbitsite}:
		
		\emph{* (star)} can substitute for exactly one word; and
		
		\emph{\# (hash)} can substitute for zero or more words.
		
	\section{APPLICATION SCENARIO: Self-organizing Streetlights}\label{section:app}
	
	In short, this experiment involves developing self-organizing streetlights. The overall goal of this application is to reduce the energy consumption while maintaining appropriate visibility in illuminated areas \cite{nascimento2017engineering}. For this purpose, each streetlight was provided with ambient brightness and motion sensors, and an actuator to control light intensity. In addition, they are able to interact with each other though an wireless communicator.


	\begin{figure}[!htb]
		\centering
		\includegraphics[height=2.2in, width=3.3in]{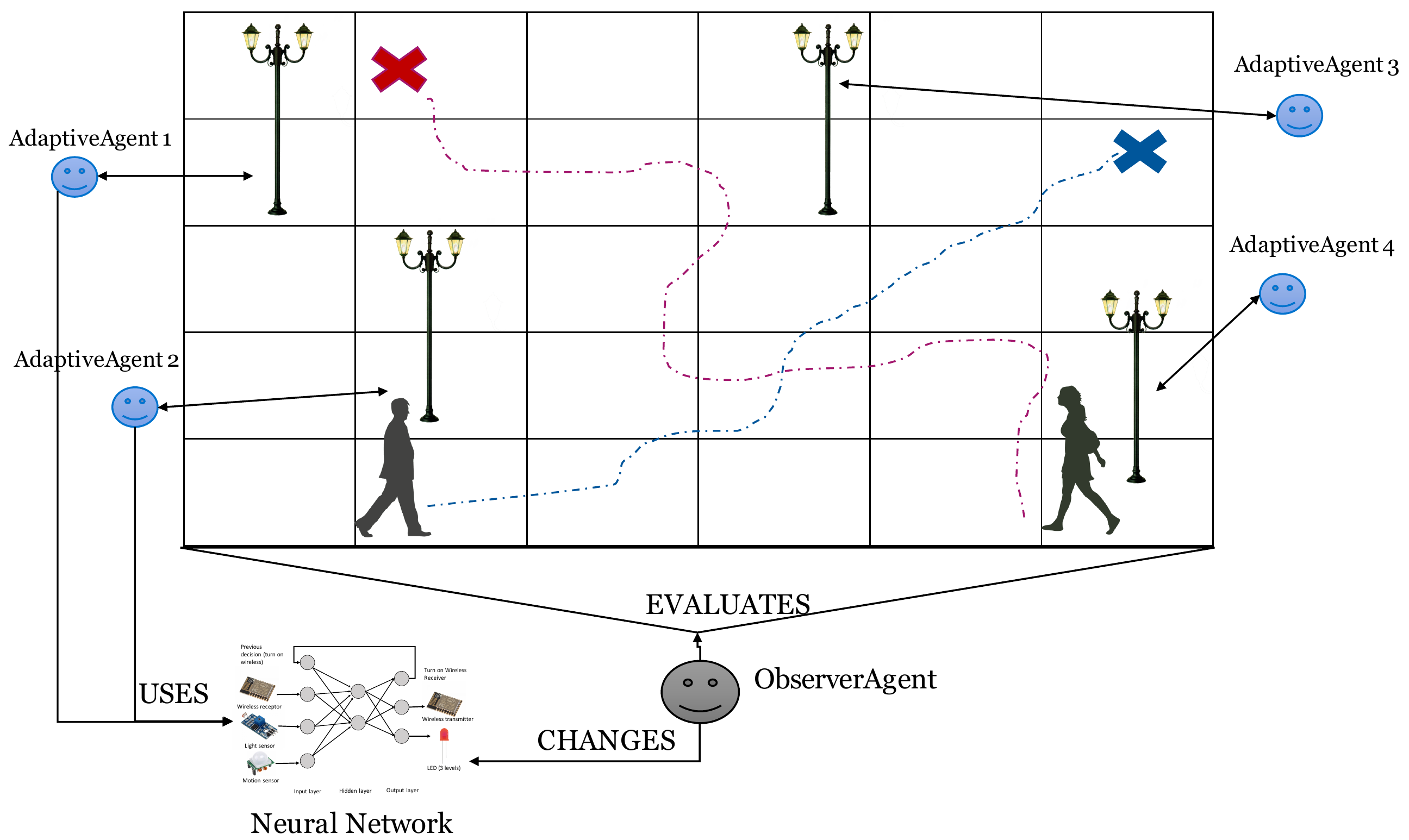}
		\caption{Overview of the general application architecture.}
		\label{figure:applicationoverview}
	\end{figure}
	
	Each street light is controlled by an AdaptiveAgent, as shown in Figure \ref{figure:applicationoverview}. We used a neuroevolutionary algorithm \cite{nascimento2017engineering} to support the design of the street behaviors of the street lights automatically. Each streetlight uses a neural network to determine the communicating signals, and whether it turns on its lights or not.
	An ObserverAgent evaluates the overall application performance and uses a genetic algorithm to optimize the AdaptiveAgents' neural network. 
	As detailed in \cite{nascimento2017engineering}, this evaluation is  based on energy consumption, the number of people that finished their routes before the simulation ends, and the total time spent by people moving during their trip:
	
	\begin{equation} 
	\begin{split}
	fitness = (1.0 \times pPeople) - (0.6 \times pTrip) -\\ 
	(0.4 \times pEnergy)
	\end{split}
	\label{eq:fitness}
	\end{equation}

\begin{figure}[!htb]
		\centering
		\includegraphics[width=8.8cm]{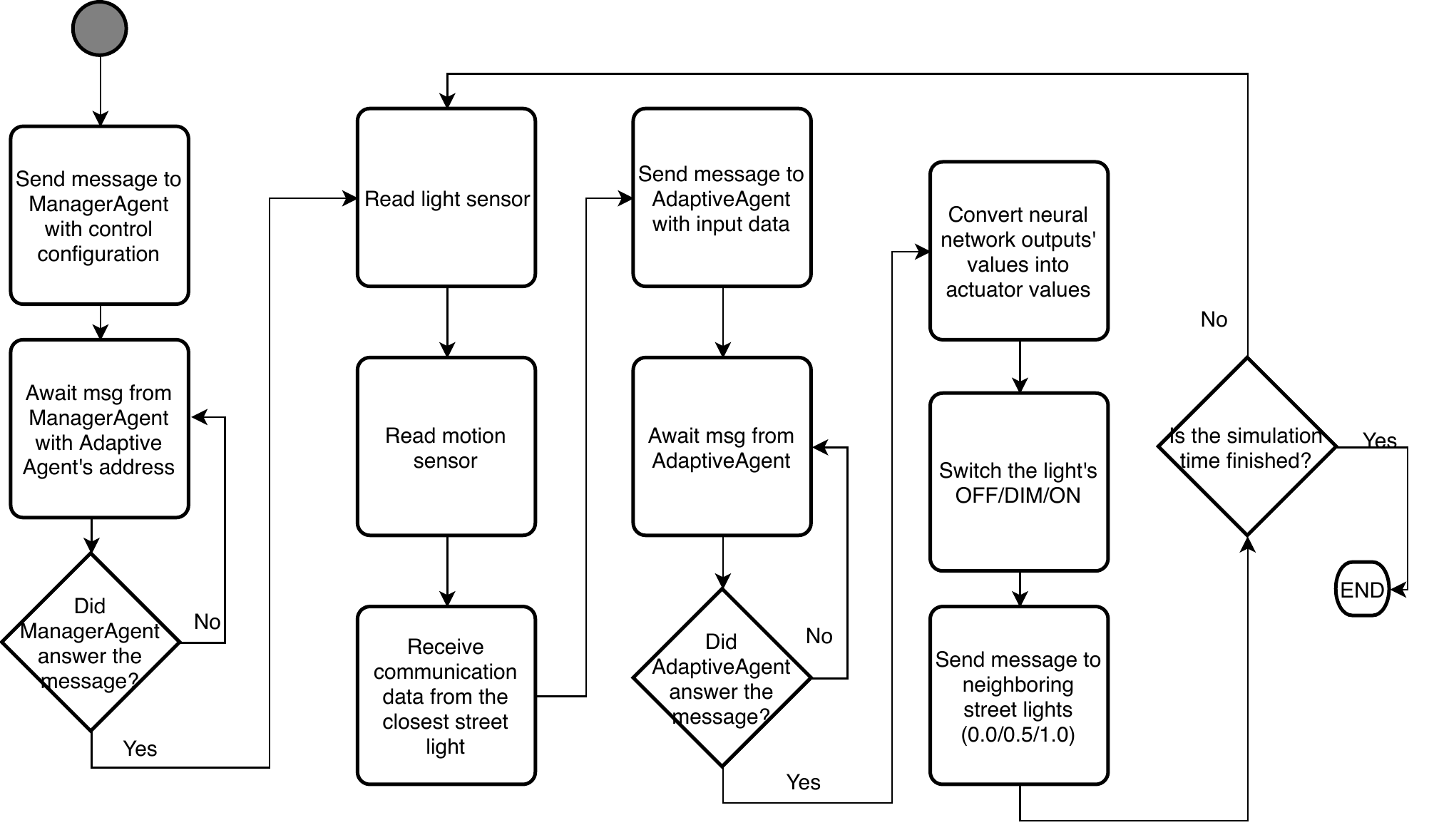}
		\caption{Activity diagram of the streetlights.}
		\label{figure:activityStreetLight}
	\end{figure}

\begin{figure}[!htb]
		\centering
		\includegraphics[width=8.8cm]{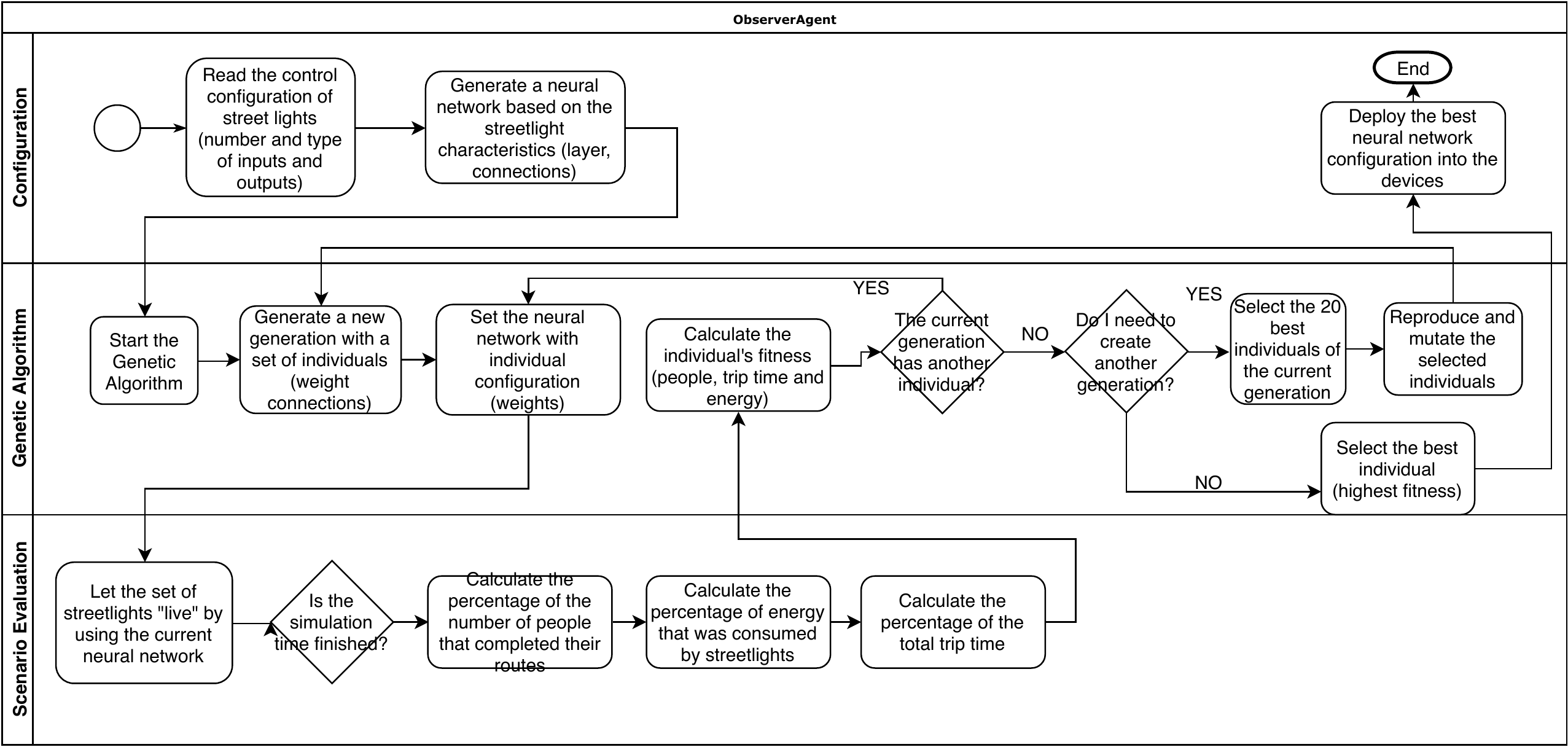}
		\caption{Activity diagram of the ObserverAgent.}
		\label{figure:activityObserverAgent}
	\end{figure}

	In order to identify the functional tests, we first created activity diagrams for the street light agents and for the ObserverAgent, as depicted in Figures \ref{figure:activityStreetLight} and \ref{figure:activityObserverAgent}. 	The interested reader may find more details about the application scenario in  \cite{nascimento2017engineering}.

	\section{Test Approach: Multilevel-based Design} \label{sec:testapproach}
	
	The main goal of a self-organizing system is to achieve global properties through local interactions. Therefore, we propose to execute several functional ad-hoc tests at local and global levels. The idea of the tests at the global level is to verify if the self-organized system solves the overall problem. If these global tests pass, we can conclude that the most basic tests (the intern ones), which were modeled at the local level, are also satisfying the functional requirements. If a global test fails, we need to understand which part of the system generated the failure, verifying the internal tests results. However, if we were executing tests at system level (performance) or evaluating how the system self-organize, we should verify the local tests independently of the global tests results. For example, according to the performance tests proposed by Kaddoum et al. \cite{kaddoum2009characterizing} to self-* systems,  we could verify whether the agents can reach the global solution by executing a desired number of steps.

\begin{figure}[!htb]
		\centering
		\includegraphics[height=2.1in, width=3.2in]{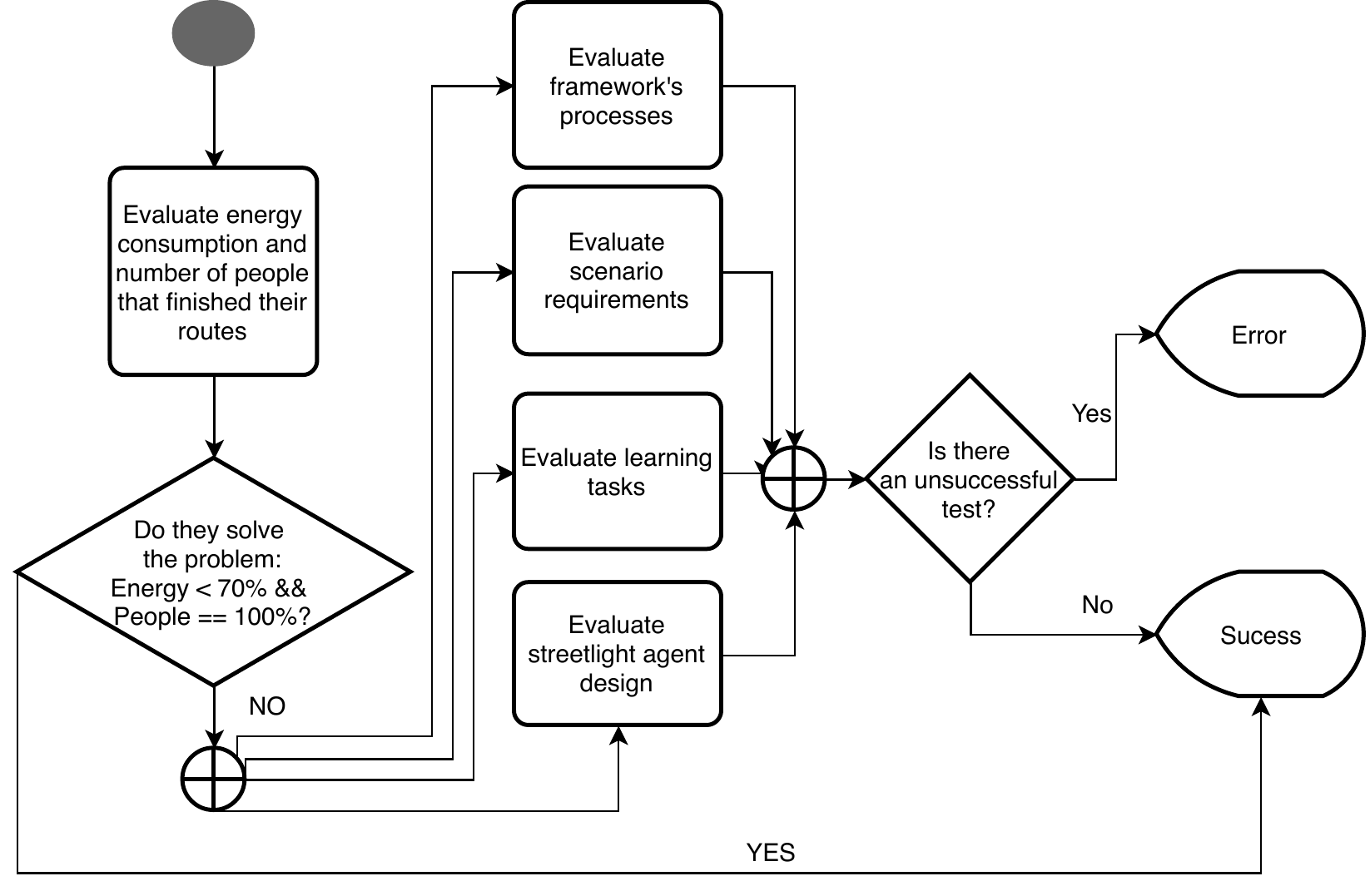}
		\caption{Testing steps.}
		\label{figure:testdiagram}
	\end{figure}

	We need to customize these tests according to the application. In general, at the global level, we should verify if the self-organized system is able to solve the problem for which it is conceived \cite{kaddoum2009characterizing}. For example, our streetlight application has the goal of achieving an specific energy consumption target and maintaining the maximum visual comfort in illuminated areas in order to enable people to finish their routes. If the multiagent system does not solve this problem, we should investigate local tasks 
	to understand why the self-organizing process failed, as depicted in Figure \ref{figure:testdiagram}. 
	
	In our illustrative example, we can investigate the failures generated by the tasks associated with the framework (i.e. the ManagerAgent cannot identify new streetlights at the scenario), to the agent design (i.e. streetlight agents must detect people, but they do not have motion sensors), tasks related to the application scenario (i.e. streetlights should communicate, but the distance between them is higher than the wireless range), or the tasks related to the learning algorithm execution (i.e. the ObserverAgent is executing the genetic algorithm wrongly, selecting the worst solutions to compose a new generation instead of the best solutions). 

	
		

	
	
	\subsection{Design and Implementation: A Publish-Subscribe based Architecture}\label{sub:TestMAS}
	
\begin{figure}[!htb]
		\centering
		\includegraphics[height=2.7in, width=3.3in]{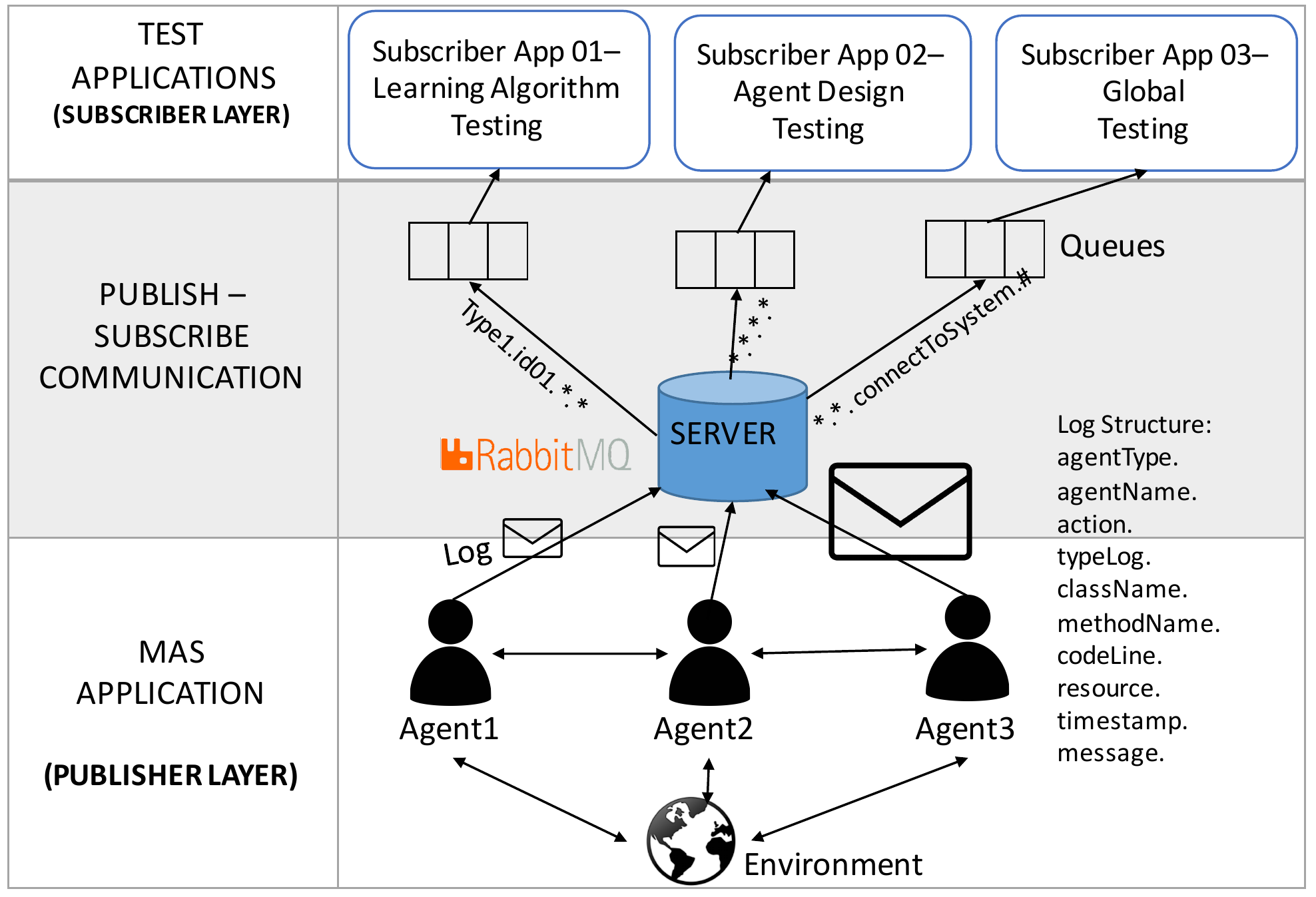}
		\caption{A Publish-Subscribe-based architecture to test MASs.}
		\label{figure:architecture}
	\end{figure}

	We developed a publish-subscribe-based architecture as a foundation for generating different kinds of test applications for MASs at different levels. Our goal is to provide mechanisms to capture and process logs generated by agents automatically. As depicted in Figure \ref{figure:architecture}, their architecture consists of three layers: MAS Application (L1), Publish-Subscribe Communication (L2), and Test Applications (L3). The Publish-Subscribe Communication layer uses the RabbitMQ platform \cite{rabbitsite} for delivering logs from agents (publishers) to be consumed by test applications (subscribers). 
	
	Each agent publishes logs with annotations that are composed of the following tags:
	\begin{itemize}
		\item  \emph{agentType}: the type of the agent (e.g OBSERVER, STREETLIGHT). In JADE, it refers to the name of the container where this agent lives;
		\item \emph{agentName}: the name provided for the agent by the system developer/user (e.g streetlight01, streetlight02, observer01);
		\item \emph{action}: the event that caused the log generation (e.g readMotionSensor, selectBestIndividuals, switchStreetLight);
		\item \emph{typeLog}: types of logs (e.g error, info, warning);
		\item	\emph{className}, \emph{methodName}, \emph{codeLine}: necessary information to identify which parts of the code generated the event;
		\item	\emph{resource}: the main resource that has been manipulated or requested by an agent during an event execution (e.g neuralController, streetlight01Info, memory). It may be used to investigate all events that are related to a specific resource;
		\item	\emph{timestamp}: time that the log was created. It is used to sort all events into a single timeline \cite{thiagopuc};
		\item	\emph{message}: a description of the event.
	\end{itemize}

	Thus, a log message must meet the pattern ``(agentType).(agentName).(action).(typeLog).
	(className).(methodName).(codeLine).(resource).(timestamp).
	(message)." Each application will have a set of values that each tag may assume, except the message tag is an open field. 

\begin{figure}[!htb]
		\centering
		\includegraphics[height=2.3in, width=3.3in]{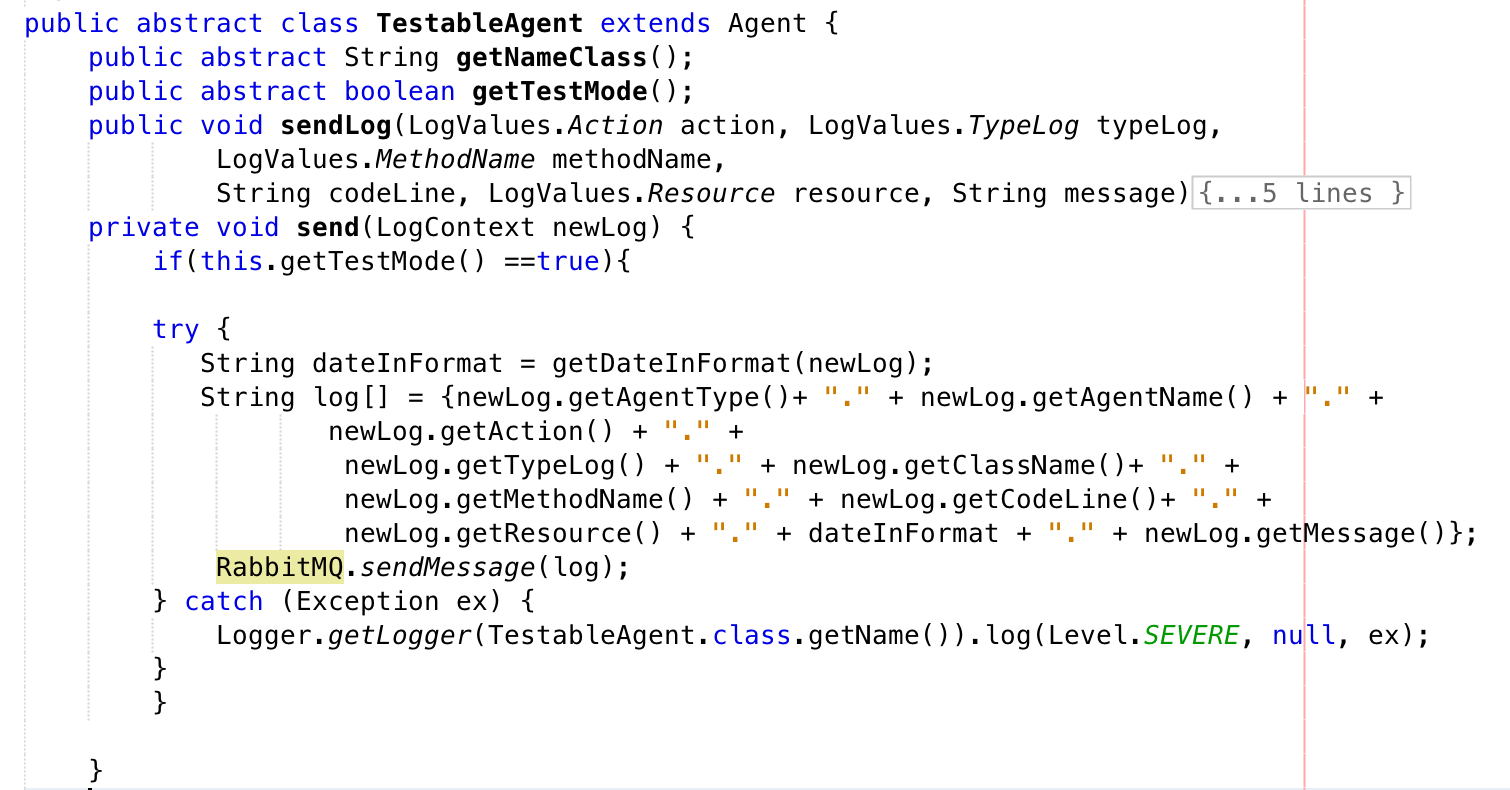}
		\caption{Testable Agent class.}
		\label{figure:testableAgentCode}
	\end{figure}

	All agents in the MAS application layer are also a TestableAgent type. As shown in Figure \ref{figure:testableAgentCode}, a Testable agent extends the JADE agent. Thus, it complies with FIPA specifications. A Testable agent uses the RabbitMQ properties to send logs with annotations as messages.

	These logs can be published from any part of the agent's code. Via the TestableAgent class and JADE properties, some tags have their values attributed autonomously, such as agentType, agentName and timestamp.  
	
	The RabbitMQ autonomously delivers log messages to queues according to their tags' values. As shown in Figure \ref{figure:architecture}, each test application defines a binding key in order to subscribe itself to consume messages from a specific queue. For example, a test application that monitors only error logs from the Observer agent must have the binding key ``Observer.*.*.error.\#." Therefore, this application will consume any log with the tuples (agentType,Observer) and (typeLog,error).  It is also possible to create applications that use multiple bindings. For example, if a performance test application needs to calculate the number of Adaptive agents that are connected to the system, this application will have to consume logs with different action values. Thus,  it needs to consume logs with the tuples (action,connectToSystem) and (action,beDestroyed).

	Test applications do not interfere on the execution of each other. Each test class extends the class RabbitMQConsumer that starts an independent process to consume messages from a specific queue.  We used the Template Method Pattern \cite{gamma1993design} to model the consumeMessage method. Thus, to consume and process particular log messages, a test class must overwrite and customize the methods getListBindingKey() and processData().

	By using queues, the publisher generates a set of information elements without the need of knowing which applications will consume them. In addition, more than one application can consume the same data, but giving them different treatments.  To understand more about the characteristics of RabbitMQ that we used in our approach, see \url{https://www.rabbitmq.com/tutorials/tutorial-five-java.html}
	(Accessed in 03/2019).
	
		\subsection{Adapting FIoT Agents to be Testable Agents}
		
		\begin{figure}[!htb]
			\centering
			\includegraphics[height=1.5in, width=2.9in]{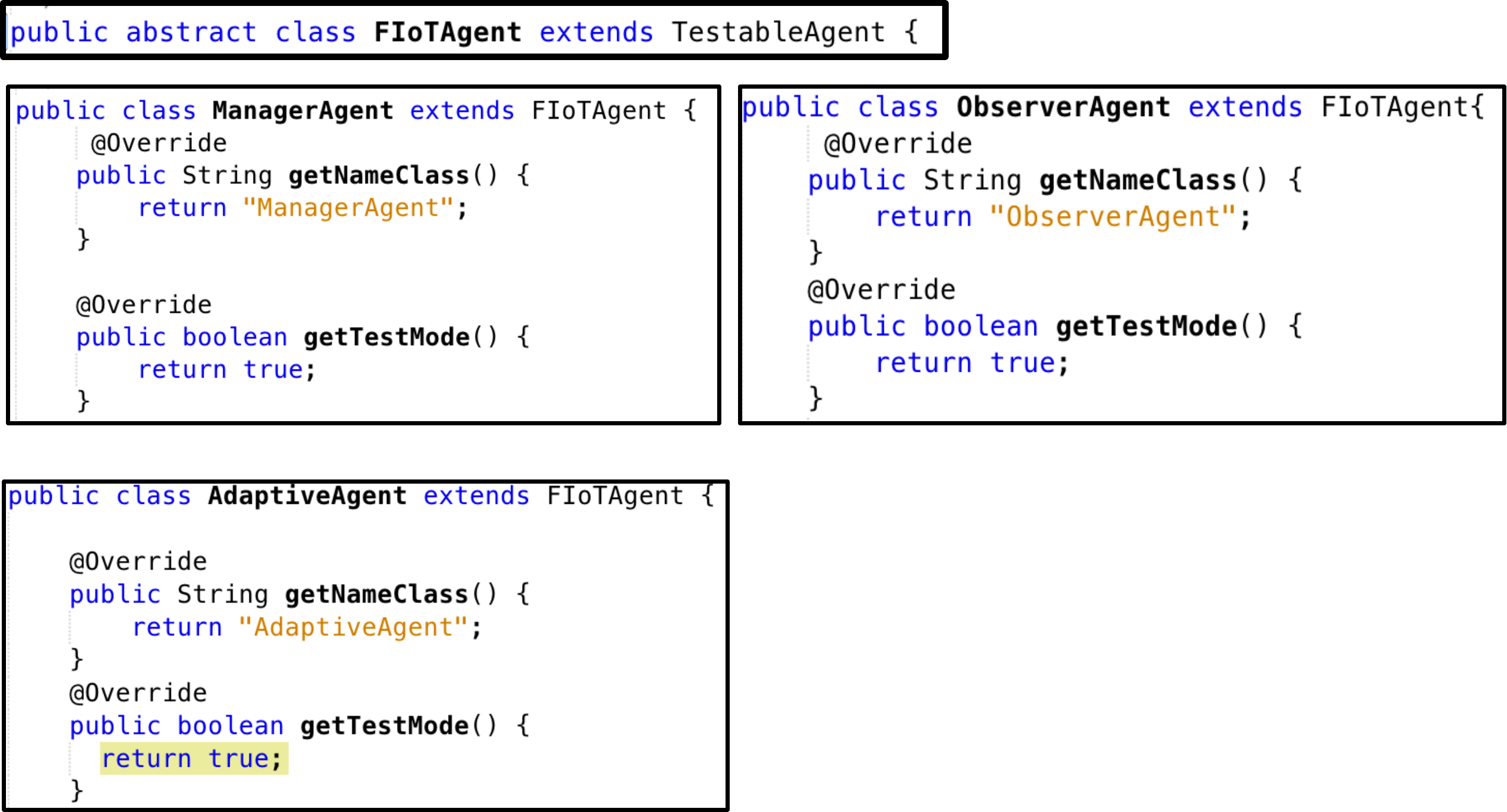}
			\caption{Making FIoT's agents as Testable Agents.}
			\label{figure:adaptingFIoTAgents}
		\end{figure}
		
		\begin{figure}[!htb]
			\centering
			\includegraphics[height=1.8in, width=3.3in]{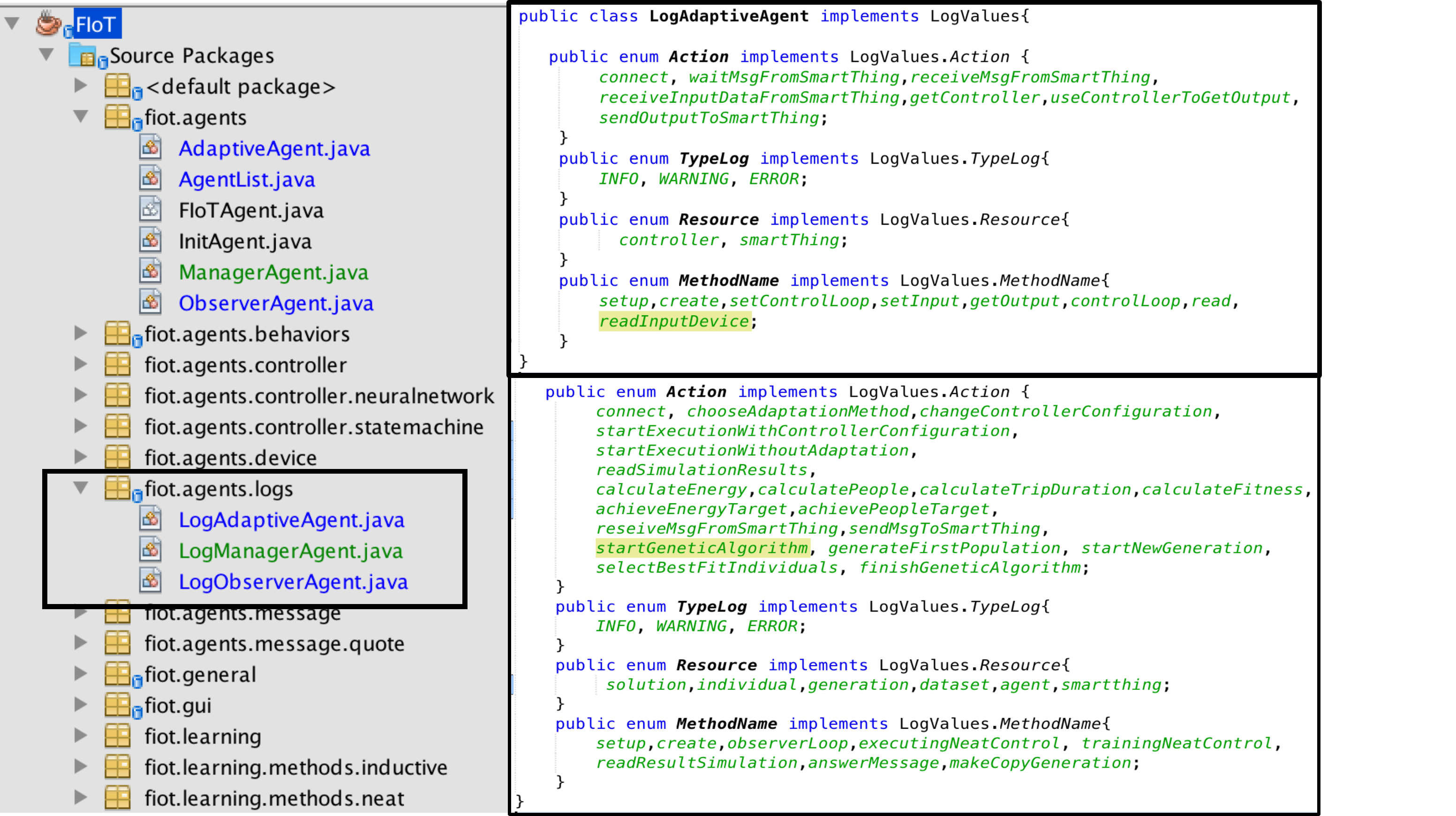}
			\caption{Setting log values for each Testable FIoT agent.}
			\label{figure:addingLogValues}
		\end{figure}
		
		Our first step was to allow FIoT agents to publish logs during the application execution, extending the TestableAgent class, as shown in Figure \ref{figure:adaptingFIoTAgents}. Then, we set the log values that can be published by each agent type. For example, the AdaptiveAgent can use the word `receiveIputDataFromSmartThing' to replace the tag action in the annotated log, while the ObserverAgent can use `startGeneticAlgorithm'.


	\section{TESTS AND RESULTS}\label{section:results}
	Our test approach takes two perspectives into account: the local and the global. The local perspective considers the tasks that an individual agent in the collection of streetlight agents must execute, such as collecting data, switching the light and communicating with the other agents. The global perspective takes the global tasks into account, such as verifying whether the self-organized system guarantees that people finish their routes before the simulation ends and whether the system achieves a pre-specified energy consumption target. 
	
	In this experiment, we have one test application consuming logs related to  the global perspective, while we have two test apps related to the local perspective:  
	one to monitor the ObserverAgent and its learning algorithm execution and another one to monitor the streetlight agents. 
	
	By using our proposed architecture, we created some test applications to execute functional tests at local and global levels. Thus, this section presents part of the test plan that we created and performed for testing the application presented in the section \ref{section:app}. 

		\subsection{Local and Global tests}
		
		We executed various test cases, taking seven parameters into account: (i) level (e.g. local or global); (ii) sub-level (e.g. related to the learning, framework, agent design or scenario requirements); (iii) function (e.g. composed of a set of actions; for example, the function evaluateSolution may be composed of the actions calculateEnergy and calculateNumberPeople); (iv) procedure (e.g. a general description of the test); (v) input (e.g. a resource, a component); (vi) expected value (e.g. the result that will be produced when executing the test if the program satisfies its intended behavior); and (vii) validation method (e.g. the strategies that a tester performs to evaluate the system, comparing the program execution against expected results). 	Each test case execution produced several logs with meta-information annotations, which were consumed by test applications. Then, we used these logs as a validation method, as shown in Table \ref{table:tests}.

		\begin{table*}[!htb]
			\centering
			\caption{Functional tests at local and global levels (Simplified Table).} 
			\label{table:tests}
\begin{tabular}{|c|c|c|c|c|c|c|}
\hline
Level                  & Sub-level & Func.                                                                                   & Procedure                                                                                                                                                                              & Input                                                                                                                  & \begin{tabular}[c]{@{}c@{}}Expected\\ value\end{tabular}                                                                                             & \begin{tabular}[c]{@{}c@{}}Validation Method\\ (logs sorted into a \\ timeline)\end{tabular}                                                                                                                                                                                                                               \\ \hline
\multirow{6}{*}{Local} & Framework & \begin{tabular}[c]{@{}c@{}}create \\ AdaptiveAgent\\ to the \\ streetlight\end{tabular} & \begin{tabular}[c]{@{}c@{}}ManagerAgent \\ creates a new \\ AdaptiveAgent\\ to the streetlight\end{tabular}                                                                            & \begin{tabular}[c]{@{}c@{}}Control \\ configuration \\ (number of \\ inputs\\ and outputs)\end{tabular}                & \begin{tabular}[c]{@{}c@{}}AdaptiveAgent\\ with the selected\\ control\end{tabular}                                                                  & \begin{tabular}[c]{@{}c@{}}1)MANAGER.\\ receiveMsgFromSmartThing.\\ *.*.*.*.smartThing.\#\\ 2)MANAGER.\\ createAdaptiveAgent.INFO.\#\\ 3)AdaptiveAgent.lightsAgent.\\ connect.\#\\  4)MANAGER.\\ sendMsgToSmartThing.INFO.\#\\ 5)AdaptiveAgent.lightsAgent.\\ receiveInputDataFromSmartThing.\#\end{tabular}               \\ \cline{2-7} 
                       & Scenario  & collect data                                                                            & \begin{tabular}[c]{@{}c@{}}streetlight 10\\ (node10) reads\\ its sensors data\end{tabular}                                                                                             & \begin{tabular}[c]{@{}c@{}}streetlight's \\ motion and \\ light sensors, \\ and \\ communication \\ input\end{tabular} & \begin{tabular}[c]{@{}c@{}}AdaptiveAgent\\ receives data \\ from the \\ streetlight's \\ sensors\end{tabular}                                        & \begin{tabular}[c]{@{}c@{}}1)lightContainer.node10.\\ receiveWirelessData.\#\\ 2)lightContainer.node10.\\ readLightSensor.\#\\ 3)lightContainer.node10.\\ readMotionSensor.\#\\ 4)lightContainer.node10.\\ sendMsg.*.*.\\ msgAdaptiveAgent\\ 5)AdaptiveAgent.lightsAgent.\\ receiveInputDataFromSmartThing.\#\end{tabular} \\ \cline{2-7} 
                       & Learning  & process output                                                                          & \begin{tabular}[c]{@{}c@{}}AdaptiveAgent \\ uses a \\ neural network\\ to process sensors\\ data and generate \\ output\end{tabular}                                                   & \begin{tabular}[c]{@{}c@{}}streetlight's \\ sensors data\end{tabular}                                                  & \begin{tabular}[c]{@{}c@{}}Adaptive agent\\ calculates \\ two outputs \\ (led and \\ wireless\\ data)\end{tabular}                                   & \begin{tabular}[c]{@{}c@{}}1)AdaptiveAgent.lightsAgent.\\ useControllerToGetOutput.\#\\ 2)AdaptiveAgent.lightsAgent.\\ sendOutputToSmartThing.\#\end{tabular}                                                                                                                                                              \\ \cline{2-7} 
                       & Scenario  & set actuators                                                                           & \begin{tabular}[c]{@{}c@{}}streetlight 10\\  sets its\\ actuators values\end{tabular}                                                                                                  & \begin{tabular}[c]{@{}c@{}}neural \\ network's \\ output values\end{tabular}                                           & \begin{tabular}[c]{@{}c@{}}Streetlight 10\\ switches its \\ light\\ and sends \\ message\\ (according to\\ the neural \\ output values)\end{tabular} & \begin{tabular}[c]{@{}c@{}}3)lightContainer.node10.\\ receiveNeuralNetworkCommand.\#\\ 4.1)lightContainer.node10.\\ switchLightON.\#\\ 4.2)lightContainer.node10.\\ switchLightOFF.\#\\ 5)lightContainer.node10.\\ sendWirelessData.\#\end{tabular}                                                                        \\ \cline{2-7} 
                       & Learning  & \begin{tabular}[c]{@{}c@{}}change the\\  neural\\ network\end{tabular}                  & \begin{tabular}[c]{@{}c@{}}ObserverAgent\\ uses an\\ individual's genes\\ to set the ANN \\ weights\\ (see \\ subsection \ref{sub:evolution})\end{tabular}                         & \begin{tabular}[c]{@{}c@{}}an individual \\ from\\ the current \\ generation\end{tabular}                              & \begin{tabular}[c]{@{}c@{}}the ANN\\ weights \\ sequence is\\ equal to the \\ current \\ individual\end{tabular}                                     & \begin{tabular}[c]{@{}c@{}}1)OBSERVER.\\ chooseAdaptationMethod.\#\\ 2)OBSERVER.\\ selectNeuralConfiguration.\#\\ 3)OBSERVER.\\ useIndividualGenesToANN.\#\\ 4)OBSERVER.\\ startExecutionWithController\\ Configuration.\#\end{tabular}                                                                                    \\ \cline{2-7} 
                       & Scenario  & \begin{tabular}[c]{@{}c@{}}switch the \\ light ON\end{tabular}                          & \begin{tabular}[c]{@{}c@{}}Streetlight Agent \\ (node 10)\\ switches the \\ light ON\end{tabular}                                                                                      & \begin{tabular}[c]{@{}c@{}}neural \\ network's \\ light output \\ value\\ is positive\end{tabular}                     & \begin{tabular}[c]{@{}c@{}}node10's \\ light sensor\\ detects a value \\ equal or \\ higher than\\ its light \\ brightness\end{tabular}              & \begin{tabular}[c]{@{}c@{}}1)lightContainer.node10.\\ receiveNeuralNetworkCommand.\#\\ 2)lightContainer.node10.\\ switchLightON.\#\\ 3)lightContainer.node10.\\ detectLight.\#\\ 4)lightContainer.lights.\\ finishSimulation.\#\end{tabular}                                                                               \\ \hline
Global                 & MAS       & \begin{tabular}[c]{@{}c@{}}evaluate the \\ selected\\  solution\end{tabular}            & \begin{tabular}[c]{@{}c@{}}Observer Agent \\ analyzes \\ the energy \\ consumption and \\ whether \\ everyone finished \\ their \\ routes during the \\ selected solution\end{tabular} & \begin{tabular}[c]{@{}c@{}}the best \\ individual\\ of the last \\ generation\end{tabular}                             & \begin{tabular}[c]{@{}c@{}}energy \\ consumption is\\ less than 70\%\\ and everybody \\ finished their \\ routes\end{tabular}                        & \begin{tabular}[c]{@{}c@{}}1)OBSERVER.\\ startExecutionWith\\ ControllerConfiguration.\#\\  2)OBSERVER.\\ readSimulationResults.\#\\ 3)OBSERVER.\\ calculateEnergy.\#\\ 4)OBSERVER.\\ achieveEnergyTarget.\#\\ 5)OBSERVER.\\ achievePeopleTarget.\#\\ 6)OBSERVER.\\ calculateFitness.\#\end{tabular}                       \\ \hline
\end{tabular}
			\end{table*}

\begin{figure}[!htb]
		\centering
		\includegraphics[width=8.9cm]{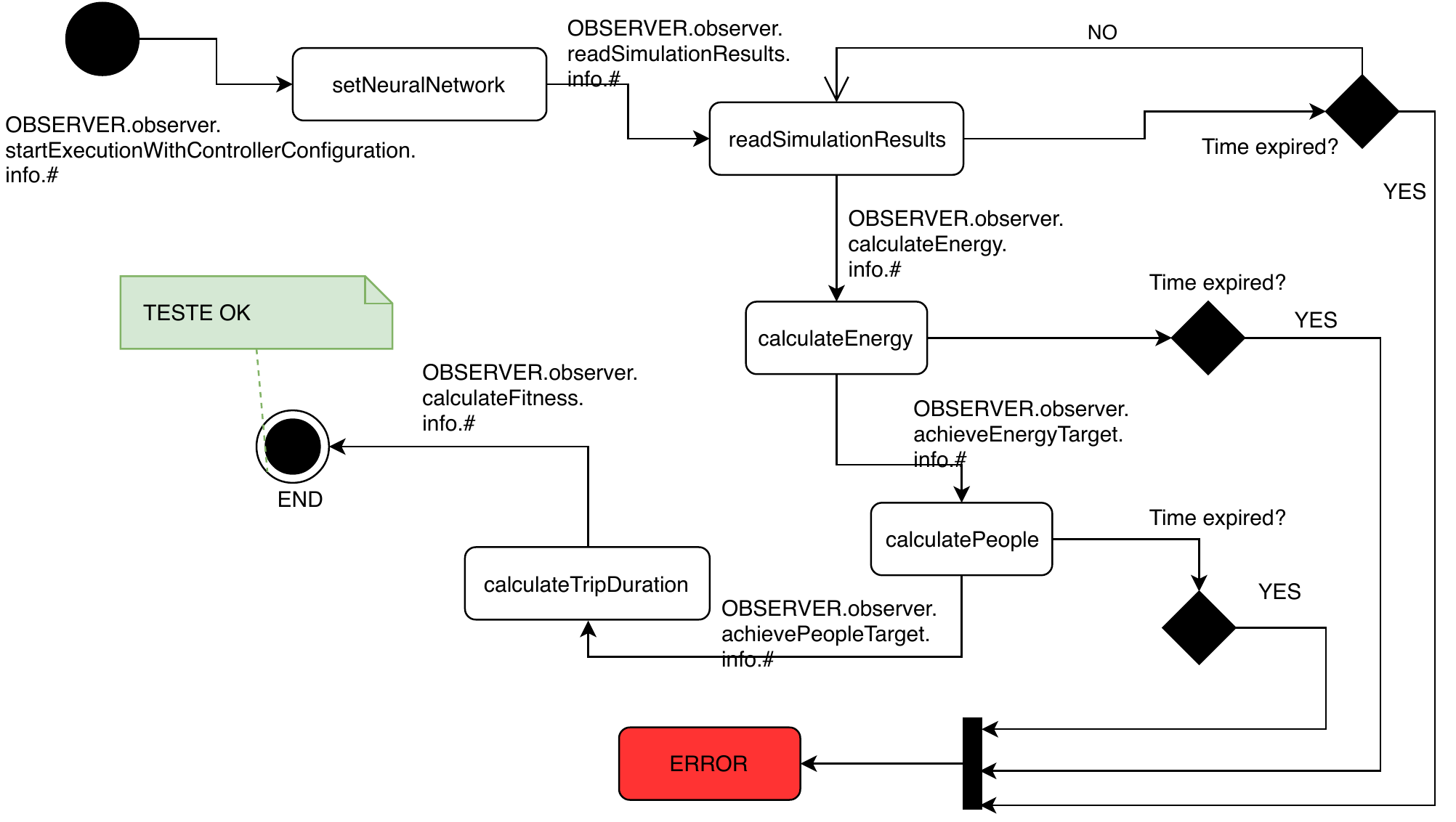}
		\caption{Simplified state machine for verifying test cases generated for the function ``evaluate selected solution".}
		\label{figure:globalstatemachine01}
	\end{figure}

\begin{figure}[!htb]
			\centering
			\includegraphics[width=8.9cm]{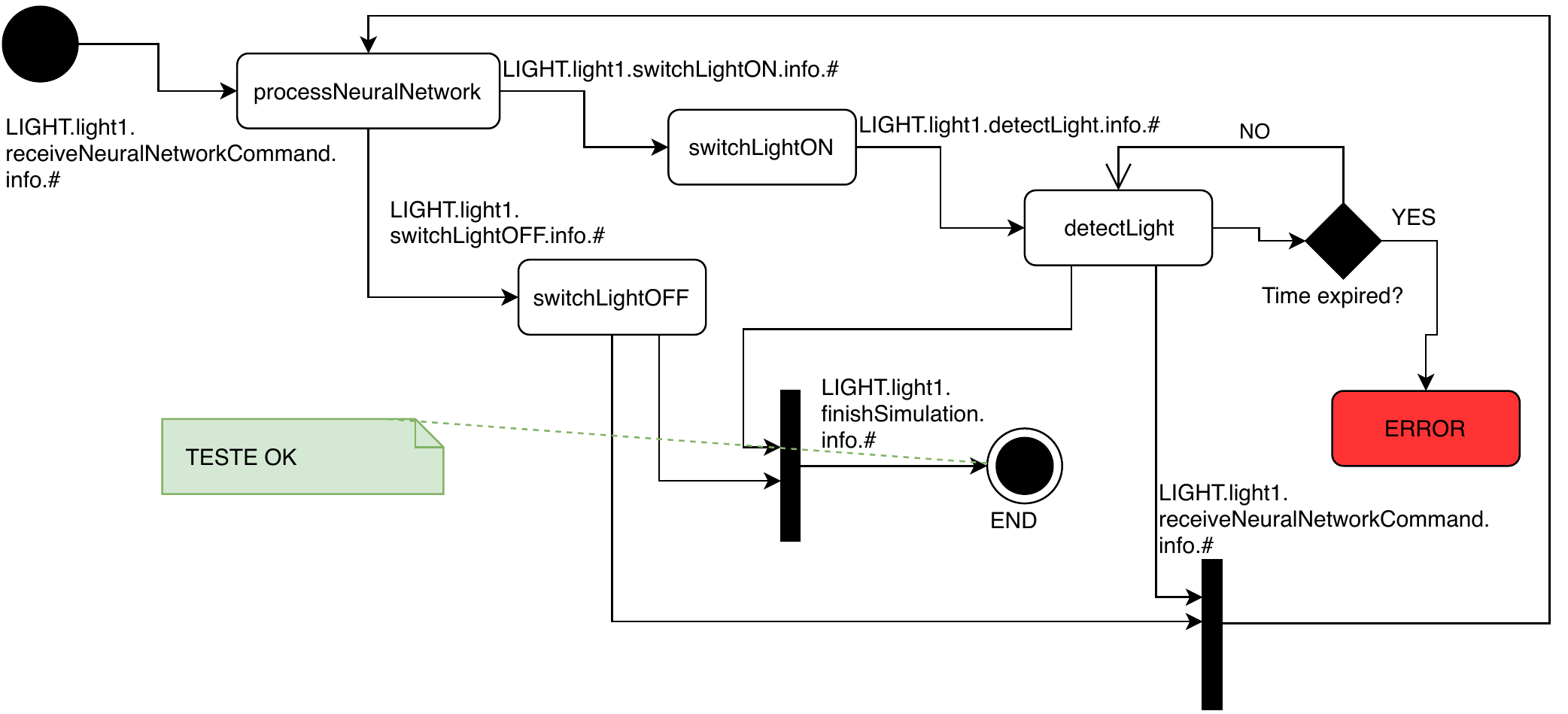}
			\caption{Simplified state machine for verifying test cases generated for the functions ``switch the light ON" and ``switch the light OFF".}
			\label{figure:statemachine01}
		\end{figure}

	To validate a test case, the test application must verify whether the logs are appearing in the order described in the Validation Method column. Therefore, after the developer informs the logs from the validation column, the test application will automatically create a state machine, where each state represents an action. For example, Figures \ref{figure:globalstatemachine01} and \ref{figure:statemachine01} illustrate the state machine that were created to validate the execution of the global test ``evaluate solution" and the local test ``switch the light ON", respectively. As shown, the verification program defines the transition between states as a log. A transition will only occur when the expected log appears. Each state has a maximum wait time for the expected log(s). Thus, if the maximum wait time exceeds a threshold, an error linked to the current state will be generated. This situation indicates that an agent performed an unexpected behavior and the action was not successful executed. For example, if the multiagent system does not self-organize to a satisfactory solution, it will not produce the log ``OBSERVER.observer.achieveEnergyTarget.\#". Thus, an error linked to the state ``calculateEnergy" will be generated, as depicted in Figure \ref{figure:globalstatemachine01}.
	
\begin{figure}[!htb]
		\centering
		\includegraphics[width=8.4cm]{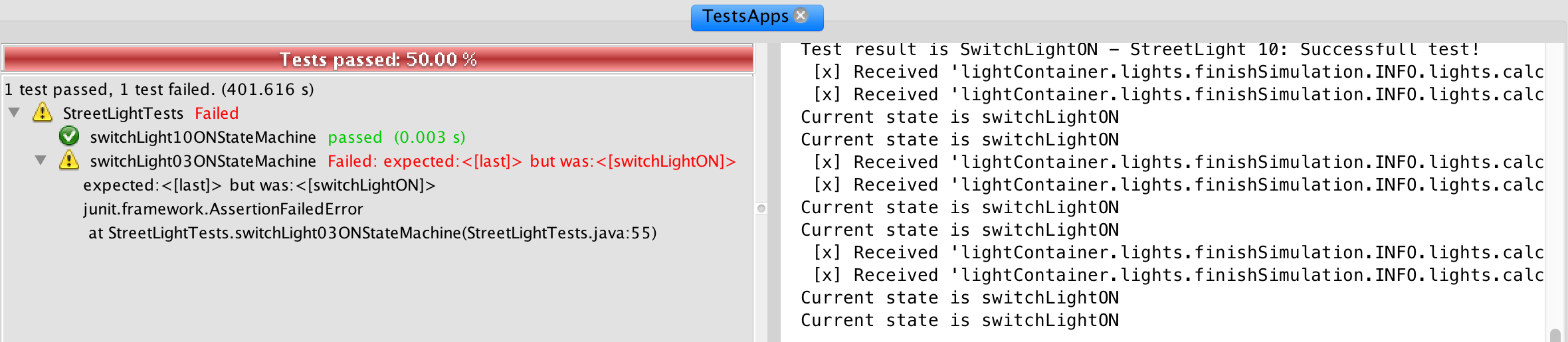}
		\caption{Executing the state machine to test the function ``switch the light ON": failure generated between states ``switchLightON" and ``detectLight" - specific log was not consumed.}
		\label{figure:localResultFail}
	\end{figure}

\begin{figure}[!htb]
		\centering
		\includegraphics[width=8.4cm]{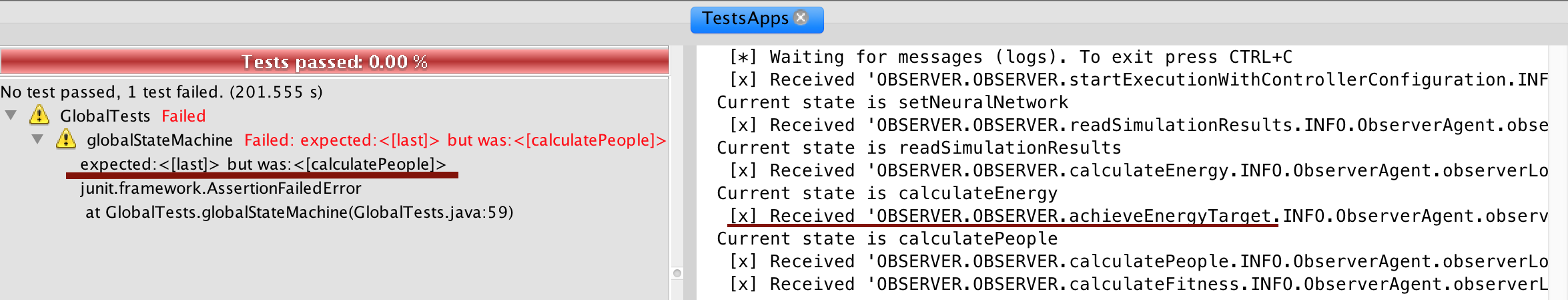}
		\caption{Executing the state machine to test the evaluation solution: failure generated between states ``calculatePeople" and ``calculateTripDuration" - because the machine did not receive the log that indicates that everyone finished their routes during the selected solution.}
		\label{figure:globalResultFail}
	\end{figure}

\begin{figure}[!htb]
		\centering
		\includegraphics[width=8.4cm]{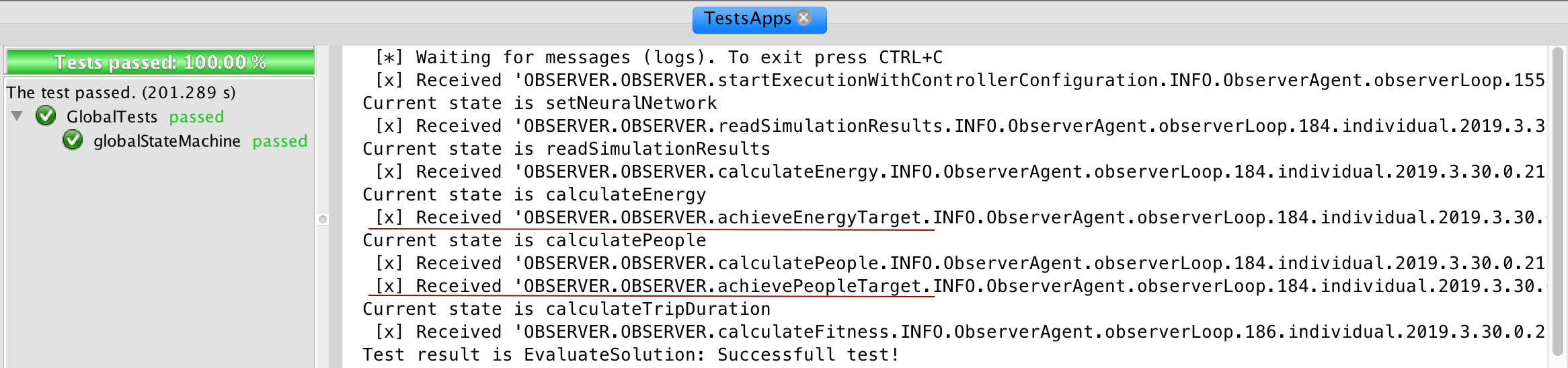}
		\caption{Executing the state machine to test the evaluation solution.}
		\label{figure:globalResultOK}
	\end{figure}

In order to force test failure and verify if these test applications were able to identify faults, we forced certain classes to act incorrectly during the execution of the program over some local tests. For example, to test the function ``switch the light ON", we inserted a defect that makes some streetlights to go dark during the simulation. 
Therefore, a streetlight agent that switched its light ON on the previous execution, did not detect brightness on the current execution and failed. As the test application did not receive the log ``LIGHT.light1.detectLight.info.\#", its state machine indicated a failure in the state ``switchLightON," as depicted in Figure \ref{figure:localResultFail}. Considering that a person can only move if his current and next positions are not completely dark, it interferes on the overall solution evaluation. Consequently, if a person does not finish his or her route, the test at the global level will also fail. 
 Figure \ref{figure:globalResultFail} depicts the logs that were generated by agents while this situation was being executed. Figure \ref{figure:globalResultOK}  depicts the global test that was executed without this defect.

\begin{figure}[!htb]
	\centering
	\includegraphics[width=8.7cm]{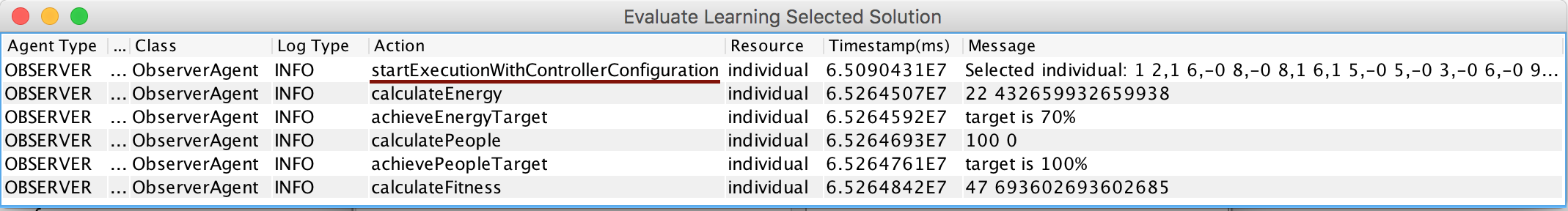}
	\caption{Subscribing only to  receive logs related to the evaluation solution testing.}
	\label{figure:logsForSolutionEvaluation}
\end{figure}

\begin{figure}[!htb]
	\centering
	\includegraphics[width=8.7cm]{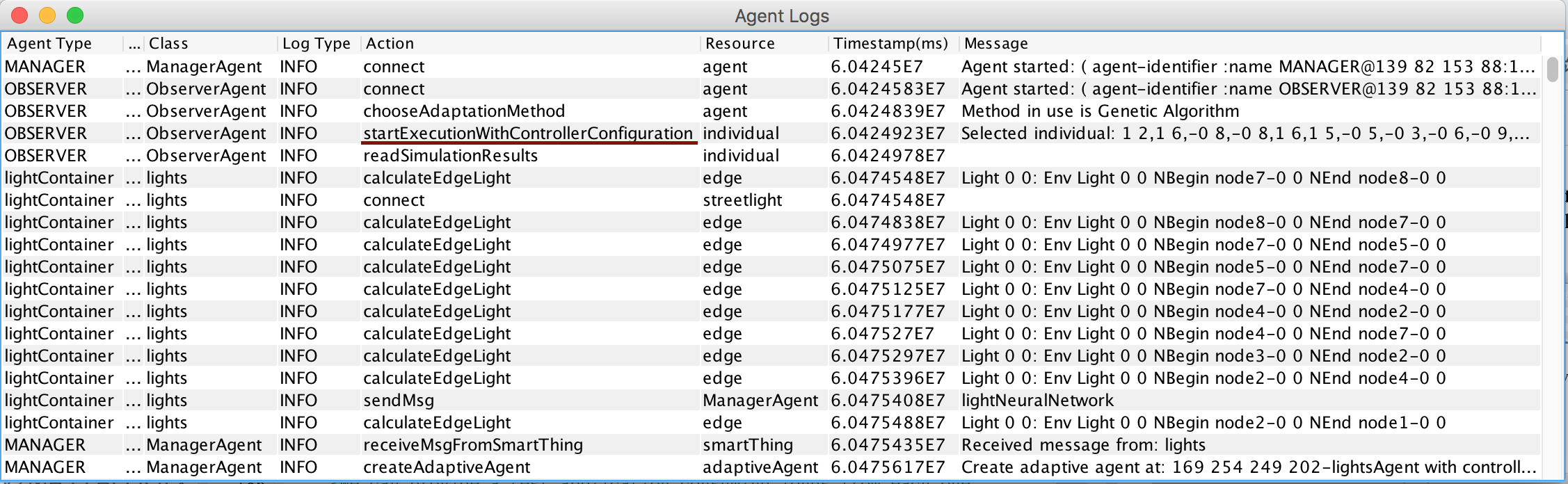}
	\caption{Subscribing to receive logs from all agents.}
	\label{figure:logsFromAllAgents}
\end{figure}

Using our proposed solution, a test application can automatically select those logs from different agents that are essential for a specific test case and present them sorted in a single timeline. As a result, the interface depicted in Figure \ref{figure:logsForSolutionEvaluation} shows just the logs that were consumed by the evaluation test application according to this binding key list. In addition, all logs are organized in a single timeline. As shown, not all logs depicted in Figure \ref{figure:logsFromAllAgents} were presented in this interface, but only the logs relevant to the execution of this test case. Thus, we were able to verify these logs in order to find the fault that generated the failure indicated by the state machine.

\subsection{Test Results}
		As shown in Table \ref{table:tests}, we executed some functional tests at local and global levels. By using state machines, the test applications were able to validate these test cases by comparing the logs consumed from the MAS publisher against the logs listed in the ``Validation Method" column. In addition, we also conducted some tests by inserting software failures and verifying if our test software could be useful for detecting these faults. As a result, after the state machine had indicated a failure, the developer could use the interface to identify the fault and reduce the diagnosis time.
	
	\section{Conclusions and Prospects}\label{section:conclusion}
	We presented a promising decoupled architecture that allows a developer to execute tests simultaneously and independently while running a MAS. In addition, we provided evidence of the applicability of our proposal, using it to test a self-organizing MAS application. We showed that it is possible to develop different tests for a self-organizing multi-agent system at local and global levels by using logs containing meta-information annotations and a publish-subscribe technology.
	
	In the following we are proposing future directions that
we intend to investigate.

		\subsection{Other Application Domains}
		In this paper, we described a self-organizing application in the IoT domain. But, our approach can also be applied to other application domains. For example, we may consider a self-organizing swarm robotics \cite{floreano2007evolutionary}, where the robot behavioral mechanisms are automatically generated by using a learning algorithm. 
		Floreano et al. \cite{floreano2007evolutionary} describes a set of robotic agents that self-organizes to forage in an environment containing a food and a poison sources. Their overall goal is to increase the robot density around the food. Thus, these robotic agents may learn to distinguish the poison source from the food source and to signal to the other robots the food position. Therefore, we could develop a test application at the global level to evaluate if all robots are at the food source after the simulation ends. At the local level, we can evaluate the learning algorithm and the physical characteristics of the robots, such as their sensors and actuators.
	\subsection{ Physical Environment }
		The self-organizing process can occur in a simulated or in a physical environment. However, many devices could be damaged if we were to use real equipment, since several configurations must be tested during the training process. Therefore, to execute the training algorithm, we decided to simulate how smart street lights behave in a fictitious neighborhood. After the training process, we can transfer the evolved neural network to physical devices and observe how they behave in a real scenario. As  our approach is based on a publish-subscribe platform, it works independent on the programming language. But we need to adapt our physical streetlights to publish logs at runtime.
	\subsection{Other perspectives}
	We considered two main perspectives: the local and the glocal. But we could have explored other perspectives, such as: (i) a framework perspective (i.e. evaluating the  agent interactions generated because of the framework that we used to create the application); (ii) a learning perspective (i.e. a test application to inspect the interactions generated because of the learning algorithms); (iii) a designing perspective (i.e. a test application to evaluate the sensors, actuators and analysis architecture that were selected to compose the agent), and (iv) a scenario perspective (i.e. a test application to consume the logs generated by the application scenario).

	\subsection{Testing Prediction and Self-Adaptive Applications}
	There are other non-deterministic characteristics that have been usually associated to current MAS systems, such as learning and self-adaptation. It is possible to extend our approach to test these kinds of applications. For example, Briot et al. \cite{briot2016multi} describe a multiagent architecture to monitor fruit storage and offer predictions about shelf life. Analogously, this application has a global goal of achieving an specific target accuracy. If this system does not present a desired result to the new dataset entries, 
	we can implement local tests to evaluate the sensors measuring the storage conditions, to test the back-propagation algorithm, and the communication among the agents.
	
	\subsection{Testing Self-organizing Neural Networks}
	According to Amari \cite{amari1990mathematical}, non supervised learning scheme is sometimes called self-organization. It occurs when a neuron modifies its weights depending only on its state and input signal, without a teacher or error signal. In such case, 
	tests at the global level may evaluate the general purpose of the self-organizing neural network, while tests at the local level may evaluate each neuron, verifying the algorithms for encoding inputs and decoding outputs, whether the input signals received by each neuron is part of the information source, whether the output of a neuron is received as an input by another neuron, etc. In addition, we can also develop a test to consume logs from the application scenario, allowing us to create a map between context changes \cite{nascimento2018context} and neural changes. 
	



\section*{Acknowledgment}
 This work has been supported by the Laboratory of Software Engineering (LES) at PUC-Rio. Our thanks to CAPES, CNPq, FAPERJ, PUC-Rio for their support through scholarships and fellowships, and the Natural Sciences and Engineering Council of Canada (NSERC).

%% file: V2MAS_PAPER.bbl
\begin{thebibliography}{10}
\providecommand{\url}[1]{#1}
\csname url@samestyle\endcsname
\providecommand{\newblock}{\relax}
\providecommand{\bibinfo}[2]{#2}
\providecommand{\BIBentrySTDinterwordspacing}{\spaceskip=0pt\relax}
\providecommand{\BIBentryALTinterwordstretchfactor}{4}
\providecommand{\BIBentryALTinterwordspacing}{\spaceskip=\fontdimen2\font plus
\BIBentryALTinterwordstretchfactor\fontdimen3\font minus
  \fontdimen4\font\relax}
\providecommand{\BIBforeignlanguage}[2]{{%
\expandafter\ifx\csname l@#1\endcsname\relax
\typeout{** WARNING: IEEEtran.bst: No hyphenation pattern has been}%
\typeout{** loaded for the language `#1'. Using the pattern for}%
\typeout{** the default language instead.}%
\else
\language=\csname l@#1\endcsname
\fi
#2}}
\providecommand{\BIBdecl}{\relax}
\BIBdecl

\bibitem{pvechouvcek2008industrial}
M.~P{\v{e}}chou{\v{c}}ek and V.~Ma{\v{r}}{\'\i}k, ``Industrial deployment of
  multi-agent technologies: review and selected case studies,''
  \emph{Autonomous Agents and Multi-Agent Systems}, vol.~17, no.~3, pp.
  397--431, 2008.

\bibitem{do2017fiot}
N.~M. do~Nascimento and C.~J.~P. de~Lucena, ``Fiot: An agent-based framework
  for self-adaptive and self-organizing applications based on the internet of
  things,'' \emph{Information Sciences}, vol. 378, pp. 161--176, 2017.

\bibitem{di2008generic}
G.~Di~Marzo~Serugendo, J.~Fitzgerald, A.~Romanovsky, and N.~Guelfi, ``A generic
  framework for the engineering of self-adaptive and self-organising systems,''
  in \emph{Dagstuhl Seminar Proceedings}.\hskip 1em plus 0.5em minus
  0.4em\relax Schloss Dagstuhl-Leibniz-Zentrum f{\"u}r Informatik, 2008.

\bibitem{gardelli2005role}
L.~Gardelli, M.~Viroli, and A.~Omicini, ``On the role of simulation in the
  engineering of self-organising systems: Detecting abnormal behaviour in
  mas,'' 2005.

\bibitem{bernon2006enhancing}
C.~Bernon, M.-P. Gleizes, and G.~Picard, ``Enhancing self-organising emergent
  systems design with simulation,'' in \emph{International Workshop on
  Engineering Societies in the Agents World}.\hskip 1em plus 0.5em minus
  0.4em\relax Springer, 2006, pp. 284--299.

\bibitem{SASO2}
O.~Babaoglu and e.~H.~Shrobe, ``Ninth ieee international conference on
  self-adaptive and self-organizing systems (saso 2015),'' IEEE, Cambridge, MA,
  USA, Tech. Rep., 21-25 September 2015.

\bibitem{nascimento2017publish}
N.~Nascimento, C.~J. Viana, A.~v. Staa, and C.~Lucena, ``A publish-subscribe
  based architecture for testing multiagent systems,'' in \emph{29th
  International Conference on Software Engineering \& Knowledge Engineering
  (SEKE'2017)}.\hskip 1em plus 0.5em minus 0.4em\relax SEKE/Knowledge Systems
  Institute, PA, USA, 2017.

\bibitem{iotVision}
J.~Gubbia, R.~Buyyab, S.~Marusic, and M.~Palaniswami, ``Internet of things
  (iot): A vision, architectural elements, and future directions,''
  \emph{Future Generation Computer Systems}, vol.~29, pp. 1645--1660, 2013.

\bibitem{nguyen2009testing}
C.~D. Nguyen, A.~Perini, C.~Bernon, J.~Pav{\'o}n, and J.~Thangarajah, ``Testing
  in multi-agent systems,'' in \emph{International Workshop on Agent-Oriented
  Software Engineering}.\hskip 1em plus 0.5em minus 0.4em\relax Springer, 2009,
  pp. 180--190.

\bibitem{serrano2012approach}
E.~Serrano, A.~Mu{\~n}oz, and J.~Botia, ``An approach to debug interactions in
  multi-agent system software tests,'' \emph{Information Sciences}, vol. 205,
  pp. 38--57, 2012.

\bibitem{botaa2004aclanalyser}
J.~Bot{\'\i}a, A.~Lopez-Acosta, and A.~Skarmeta, ``Aclanalyser: A tool for
  debugging multi-agent systems,'' 2004.

\bibitem{fipa}
FIPA, ``The foundation for intelligent physical agents,'' http://www.fipa.org/,
  10 2016.

\bibitem{malkomes2017cooperative}
G.~Malkomes, K.~Lu, B.~Hoffman, R.~Garnett, B.~Moseley, and R.~Mann,
  ``Cooperative set function optimization without communication or
  coordination,'' in \emph{Proceedings of the 16th International Conference on
  Autonomous Agents and Multiagent Systems}, 2017.

\bibitem{kaddoum2009characterizing}
E.~Kaddoum, M.-P. Gleizes, J.-P. Georg{\'e}, and G.~Picard, ``Characterizing
  and evaluating problem solving self-* systems,'' in \emph{2009 Computation
  World: Future Computing, Service Computation, Cognitive, Adaptive, Content,
  Patterns}.\hskip 1em plus 0.5em minus 0.4em\relax IEEE, 2009, pp. 137--145.

\bibitem{kaddoum2010criteria}
E.~Kaddoum, C.~Raibulet, J.-P. Georg{\'e}, G.~Picard, and M.-P. Gleizes,
  ``Criteria for the evaluation of self-* systems,'' in \emph{Proceedings of
  the 2010 ICSE Workshop on Software Engineering for Adaptive and Self-Managing
  Systems}.\hskip 1em plus 0.5em minus 0.4em\relax ACM, 2010, pp. 29--38.

\bibitem{nathalia:mestrado:15}
N.~M. Nascimento, ``{FIoT}: An agent-based framework for self-adaptive and
  self-organizing internet of things applications,'' Master's thesis, PUC-Rio,
  Rio de Janeiro, Brazil, August 2015.

\bibitem{trianni2011engineering}
V.~Trianni and S.~Nolfi, ``Engineering the evolution of self-organizing
  behaviors in swarm robotics: A case study,'' \emph{Artificial life}, vol.~17,
  no.~3, pp. 183--202, 2011.

\bibitem{floreano2007evolutionary}
D.~Floreano, S.~Mitri, S.~Magnenat, and L.~Keller, ``Evolutionary conditions
  for the emergence of communication in robots,'' \emph{Current biology},
  vol.~17, no.~6, pp. 514--519, 2007.

\bibitem{tanese1989distributed}
R.~Tanese, ``Distributed genetic algorithms for function optimization,'' 1989.

\bibitem{Nolfi2016}
S.~Nolfi, J.~Bongard, P.~Husbands, and D.~Floreano, \emph{Evolutionary
  Robotics}.\hskip 1em plus 0.5em minus 0.4em\relax Cham: Springer
  International Publishing, 2016, ch.~76, pp. 2035--2068.

\bibitem{thiagopuc}
T.~P. de~Ara{\'u}jo and A.~von Staa, ``Supporting failure diagnosis with logs
  containing meta-information annotations,'' \emph{Technical Reports in
  Computer Science. PUC-Rio. ISSN 0103-9741}, vol.~14, p.~21, 2014.

\bibitem{rabbitsite}
RabbitMQ, ``Rabbitmq,'' Available in https://www.rabbitmq.com/, 10 2016.

\bibitem{nascimento2017engineering}
N.~M. NASCIMENTO and C.~J.~P. LUCENA, ``Engineering cooperative smart things
  based on embodied cognition,'' in \emph{NASA/ESA Conference on Adaptive
  Hardware and Systems (AHS 2017)}.\hskip 1em plus 0.5em minus 0.4em\relax
  IEEE, 2017.

\bibitem{gamma1993design}
E.~Gamma, R.~Helm, R.~Johnson, and J.~Vlissides, ``Design patterns: Abstraction
  and reuse of object-oriented design,'' in \emph{European Conference on
  Object-Oriented Programming}.\hskip 1em plus 0.5em minus 0.4em\relax
  Springer, 1993, pp. 406--431.

\bibitem{briot2016multi}
J.-P. Briot, N.~M. de~Nascimento, and C.~J.~P. de~Lucena, ``A multi-agent
  architecture for quantified fruits: Design and experience,'' in \emph{28th
  International Conference on Software Engineering \& Knowledge Engineering
  (SEKE'2016)}.\hskip 1em plus 0.5em minus 0.4em\relax SEKE/Knowledge Systems
  Institute, PA, USA, 2016.

\bibitem{amari1990mathematical}
S.-I. Amari, ``Mathematical foundations of neurocomputing,'' \emph{Proceedings
  of the IEEE}, vol.~78, no.~9, pp. 1443--1463, 1990.

\bibitem{nascimento2018context}
N.~Nascimento, P.~Alencar, C.~Lucena, and D.~Cowan, ``A context-aware machine
  learning-based approach,'' in \emph{Proceedings of the 28th Annual
  International Conference on Computer Science and Software Engineering}.\hskip
  1em plus 0.5em minus 0.4em\relax IBM Corp., 2018, pp. 40--47.

\end{thebibliography}
